\documentstyle{mn}

\newif\ifAMStwofonts
\def\lax    {${_<\atop^{\sim}}$}
\def\gax    {${_>\atop^{\sim}}$}

\def\etal   {{\it et al.}~}

\ifoldfss
  \ifCUPmtlplainloaded \else
    \NewTextAlphabet{textbfit} {cmbxti10} {}
    \NewTextAlphabet{textbfss} {cmssbx10} {}
    \NewMathAlphabet{mathbfit} {cmbxti10} {} % for math mode
    \NewMathAlphabet{mathbfss} {cmssbx10} {} %  "   "    "
  \fi
  \ifAMStwofonts
    \ifCUPmtlplainloaded \else
      \NewSymbolFont{upmath} {eurm10}
      \NewSymbolFont{AMSa} {msam10}
      \NewMathSymbol{\upi}     {0}{upmath}{19}
      \NewMathSymbol{\umu}     {0}{upmath}{16}
      \NewMathSymbol{\upartial}{0}{upmath}{40}
      \NewMathSymbol{\leqslant}{3}{AMSa}{36}
      \NewMathSymbol{\geqslant}{3}{AMSa}{3E}

    \fi
  \fi
\fi % End of OFSS

\ifnfssone
  \newmathalphabet{\mathit}
  \addtoversion{normal}{\mathit}{cmr}{m}{it}
  \addtoversion{bold}{\mathit}{cmr}{bx}{it}
  \newmathalphabet{\mathbfit} % math mode version of \textbfit{..}
  \addtoversion{normal}{\mathbfit}{cmr}{bx}{it}
  \addtoversion{bold}{\mathbfit}{cmr}{bx}{it}
  \newmathalphabet{\mathbfss} % math mode version of \textbfss{..}
  \addtoversion{normal}{\mathbfss}{cmss}{bx}{n}
  \addtoversion{bold}{\mathbfss}{cmss}{bx}{n}
  \ifAMStwofonts
    \ifCUPmtlplainloaded \else
      \UseAMStwoboldmath
      \makeatletter
      \new@mathgroup\upmath@group
      \define@mathgroup\mv@normal\upmath@group{eur}{m}{n}
      \define@mathgroup\mv@bold\upmath@group{eur}{b}{n}
      \edef\UPM{\hexnumber\upmath@group}
      \new@mathgroup\amsa@group
      \define@mathgroup\mv@normal\amsa@group{msa}{m}{n}
      \define@mathgroup\mv@bold\amsa@group{msa}{m}{n}
      \edef\AMSa{\hexnumber\amsa@group}
      \makeatother
      \mathchardef\upi="0\UPM19
      \mathchardef\umu="0\UPM16
      \mathchardef\upartial="0\UPM40
      \mathchardef\leqslant="3\AMSa36
      \mathchardef\geqslant="3\AMSa3E
    \fi
  \fi
\fi % End of NFSS release 1

\ifnfsstwo
  \DeclareMathAlphabet{\mathbfit}{OT1}{cmr}{bx}{it}
  \SetMathAlphabet\mathbfit{bold}{OT1}{cmr}{bx}{it}
  \DeclareMathAlphabet{\mathbfss}{OT1}{cmss}{bx}{n}
  \SetMathAlphabet\mathbfss{bold}{OT1}{cmss}{bx}{n}
  \ifAMStwofonts
    \ifCUPmtlplainloaded \else
      \DeclareSymbolFont{UPM}{U}{eur}{m}{n}
      \SetSymbolFont{UPM}{bold}{U}{eur}{b}{n}
      \DeclareSymbolFont{AMSa}{U}{msa}{m}{n}
      \DeclareMathSymbol{\upi}{0}{UPM}{"19}
      \DeclareMathSymbol{\umu}{0}{UPM}{"16}
      \DeclareMathSymbol{\upartial}{0}{UPM}{"40}
      \DeclareMathSymbol{\leqslant}{3}{AMSa}{"36}
      \DeclareMathSymbol{\geqslant}{3}{AMSa}{"3E}
    \fi
  \fi
\fi % End of NFSS release 2

\ifCUPmtlplainloaded \else
  \ifAMStwofonts \else % If no AMS fonts
    \def\upi{\pi}
    \def\umu{\mu}
    \def\upartial{\partial}
  \fi
\fi

\title{Narrow Line Seyfert 1 Galaxies and the Evolution of Galaxies \& Active
Galaxies}
\author[Smita Mathur]
       {Smita Mathur \\
        Astronomy Department, The Ohio State Univ., Columbus, OH 43210}
\pagerange{\pageref{firstpage}--\pageref{lastpage}}
\pubyear{2000}

\begin{document}

\maketitle

\label{firstpage}

\begin{abstract}

 Narrow Line Seyfert 1 galaxies (NLS1s) are intriguing due to their
 continuum as well as emission line properties.  The observed peculiar
 properties of the NLS1s are believed to be due to accretion rate
 close to Eddington limit. As a consequence, for a given luminosity,
 NLS1s have smaller black hole (BH) masses compared to normal Seyfert
 galaxies. Here we argue that NLS1s might be Seyfert galaxies in their
 early stage of evolution and as such may be low redshift, low
 luminosity analogues of high redshift quasars. We propose that NLS1s
 may reside in rejuvenated, gas rich galaxies. We also argue in favor of
 collisional ionization for production of FeII in active galactic
 nuclei (AGN).

\end{abstract}


\section{Introduction}

 Years after their discovery (Osterbrock \& Pogge 1985) the NLS1s have
attracted attention of the AGN community at least partly due to their
peculiar X-ray properties.  The NLS1s are Seyfert 1 galaxies with
relatively narrow widths of permitted optical emission lines (full
width at half maximum (FWHM)\lax 2000 km/s), strong optical
FeII/H$\beta$ ratio, weak [OIII] emission, and as such were found to
occupy one extreme end of the Boroson \& Green (1992) ``eigenvector
1''. An extremely strong anti-correlation was found between soft X-ray
spectral slopes and H$\beta$ FWHM in Seyfert 1s (Boller, Brandt \&
Fink 1996) and quasars (Laor \etal 1997) meaning that a relation
between the lines and continuum exists.  Eigenvector 1 was later found
(Brandt \& Boller 1998) to correlate strongly with the soft X-ray
power-law slope. Since the soft X-rays are formed in the vicinity of
the central black hole, eigenvector 1 has probably a more fundamental
physical meaning.

NLS1s, as a class, show peculiar continuum properties as well. They
have very steep soft X-ray slopes ($<\alpha>=2.13, F_{\nu}\propto
\nu^{-\alpha}$, while for ``normal'' Sy1s $<\alpha>=1.34\pm0.03$) and
sometimes show rapid large amplitude variability.  The hard X-ray
spectra are steep as well (Brandt, Mathur \& Elvis 1997).  In the
optical and UV range, most of NLS1s show a weak ``big blue bump''
(BBB), which is most likely due to the shift of the BBB (sometimes out
of the optical/UV range) towards higher energies. The high energy tail
of the BBB is apparent as the unusually strong and steep soft X-ray
excess. NLS1s also show strong IR emission and some high polarization.

Many continuum properties of the NLS1s can be explained in terms of
high accretion rate compared to the Eddington limit (\.m =
\.M/\.M$_{Edd}$) and so, a small black hole (BH) mass for a given
luminosity. The high accretion rate explanation for the X-ray
properties of NLS1s was first proposed by Pounds, Done \& Osborne
(1995), in analogy with Galactic BH candidates whose soft X-ray
spectra become steep in their high state.  The narrow widths of the
emission lines can be explained if the Broad Line Region (BLR) scales
as L$^{1/2}$ and emission line clouds are virialized around the small
mass BH (Laor \etal 1997). As an alternative, Wandel (1997) has argued
that the continuum, with steeper X-ray slope, has stronger ionizing
power, and hence the BLR is formed at a larger distance from the
center. The resulting smaller velocity dispersion produces narrower
lines. In general there is a reasonable consensus that large \.m is
the cause of the observed peculiar properties of NLS1 (an alternative
being a pole-on view). A natural question to ask as a next step would
be ``what determines the accretion rate in an active galaxy?'' Is it the age?

%\section[]{Evolution of Galaxies and Quasars}

\section[]{Are NLS1s the Active Galaxies in the making?}

 Here we present a number of arguments in support of our proposal that
NLS1 might be Active Galaxies in early phase of their evolution. \\
1). Smaller BH mass. As per the well known correlation of Magorrian
\etal~(1998) smaller mass BHs reside in galaxies with smaller
spheroids. Since NLS1s have relatively smaller mass BHs compared to
normal Seyferts, the spheroids of their host galaxies might be smaller
(see also Laor, 1998). Indeed, in the compilation of Wandel (1999),
the NLS1 galaxy NGC4051 has the smallest black hole to bulge mass
ratio. An accreting BH would also grow in mass with time [the Salpeter
time scale of growth is determined by t$_s = 3 \times 10^7
(L_{Edd}/L_B) \eta_{0.1}$ yr. where $\eta_{0.1}$ is the radiative
efficiency in the units of 0.1 (see Fabian 1999)]. Since NLS1 accrete
at close to Eddington limit, their BHs would grow faster. So, smaller
BHs in NLS1s are likely to be younger as well.

%2). Larger accretion rate. Young, gas rich galaxies would have
%  reservoir of gas to sustain the larger accretion rate of NLS1.

2). Super-solar gas phase metallicities. There are a couple of lines
  of evidence to suggest that NLS1s may have super-solar gas phase
  metallicities. One comes from the study of high ionization emission
  lines. Wills \etal~(1999) found that the strength of  NV $\lambda
  1240$ emission line was systematically larger while the strength of
  the CIV $\lambda 1549$ was systematically smaller in AGN with narrow
  emission lines. The NV/CIV ratio serves as an abundance indicator as
  shown by Hamann \& Ferland (1993). So, the NLS1s may have large
  nitrogen abundance. Absorption lines, being insensitive to density,
  serve as better indicators of metallicities. In the narrow line AGN
  PG1404+226, Ulrich \etal~(1999) found that while the strengths of
  Ly$\alpha$ and CIV absorption lines were in reasonable agreement
  with those expected from the ionized X-ray absorber (See Mathur 1997
  for X/UV absorber models), the NV absorption line was
  significantly stronger. This observation, again, can be understood
  in terms of high nitrogen abundance in this narrow line AGN (Mathur
  \& Komossa, 2000).

  Large nitrogen abundance is obtained when overall metallicities are
  high with N$/$H $\propto$ (Z$/$Z$_{\odot}$)$^2$ where Z$_{\odot}$
  is the solar abundance. Nitrogen is preferentially enhanced because
  of secondary CNO nucleosynthesis (see Hamann \& Ferland 1999, HF99,
  for details on AGN metallicities). Thus the observations of emission
  as well as absorption lines in NLS1s imply super-solar gas phase
  metallicities. The strength of the fluorescent Fe-K alpha line in
  some NLS1s is also indicative of super-solar abundance (Fabian 1999).

  Such metal enrichment is possible when the initial mass function of
  star formation is flat, favorable for high mass star formation, and
  the evolution is fast. Such a star formation scenario is likely to
  be present in deep potential wells like galactic nuclei and
  protogalactic clumps (HF99). Moreover, high metallicities are
  achieved while consuming less gas (HF99). The NLS1s may then
  represent that early phase in galactic evolution when rapid star
  formation is taking place in the nucleus.

3). IR brightness. Many NLS1s are observed to be bright in the
  infra-red (Moran, Halpern, \& Helfand 1996). Young, star forming
  galaxies are also bright IR sources. It is possible that a part of
  the nuclear IR flux is from a nuclear star burst.

4) Analogy with high redshift quasars. In $\S 3$ we argue that NLS1s
   are analogous to very high redshift quasars. High redshift (z$>$4) quasars
   are believed to quasars in the early phase of evolution compared to
   the $z\approx 1$ quasars. By analogy, NLS1s may as well be in the
   early evolutionary phase compared to the normal Seyfert
   galaxies.\footnote{We note that Grupe (1996) has also argued that
   the age since an AGN was born might be the underlying reason for
   some NLS1-type correlations he has studied.}

\section[]{Are NLS1s low redshift analogues of high redshift quasars?}

 That the AGN phenomenon was so much stronger at z$\sim$ 2--3 than
 today has long elicited the suspicion that there is a connection
 between the youth of a galaxy and the likelihood that an AGN forms
 inside it. The question then naturally arises, ``what are the local
 counterparts to the young galaxies in early universe, in which
 local AGN may live?'' (see, e.g, Krolik 1999). A standard answer to
 this question is ``Starburst galaxies''. Heckman (1999) has argued
 that starburst galaxies are the low redshift analogues of Lyman break
 galaxies at high redshift. Similarly, we ask, what are the low
 redshift analogues of high redshift (z\gax 4) quasars? We propose
 that they might be NLS1s.

 It is interesting to note the similarity of the properties of NLS1s
 with high redshift (z\gax 4) quasars.

1) Hamann \& Ferland (1993) found high
 metallicities in high redshift quasars (Z\gax Z$_{\odot}$ at z\gax
 4). Large metallicities in NLS1s may make them low redshift, low
 luminosity analogues of high redshift quasars.

2). NLS1s and BALQSOs. Similarities between the observed properties of
  low ionization Broad Absorption Lines Quasars (BALQSOs) and NLS1s
  have been reported in the literature (e.g. Lawrence \etal~1997,
  Leighly \etal~1997). Both these classes show strong FeII$\lambda
  4570$ and AlIII$\lambda 1857$ and weak CIV$\lambda 1549$ and
  [OIII]$5007$ emission lines. Their continua are red in the optical
  and strong in the IR. Evidence of relativistic outflow is also
  reported in three NLS1s (Leighly \etal~1997). If these two classes
  are indeed related (Brandt 1999), then NLS1, at least those those
  with some evidence of outflow, might be low redshift, low luminosity
  cousins of BALQSOs. BALQSOs are tentatively identified with that
  phase in quasar evolution when the matter around the nuclear BH is
  being blown away, and a quasar emerges (see, e.g. Fabian 1999). NLS1s may
  then represent a similar early evolutionary phase at low redshift.

3). Optical spectra of a sample of
  z\gax 4 quasars revealed that their emission lines are typically
  narrower than the low redshift quasars (FWHM \lax 2000 km/s, Shields
  \& Hamann 1997). The normal explanation of this observation is that
  these are type 2 quasars, where the broad emission lines are
  obscured from our line of sight. Alternatively, these high z quasars
  might be true ``narrow'' broad line objects.

4) We note here another interesting connection with high redshift. As
 discussed in $\S 1$, NLS1s have strong FeII emission lines.  Quasars
 Q0014+813 and Q0636+680 at redshifts z=3.398 and z=3.195 respectively,
 were observed to have very strong FeII emission (Elston, Thompson, \&
 Hill 1994). Are they also highly accreting objects at early
 evolutionary phase? Note also the narrow UV emission lines (FWHM \lax
 2150 km s$^{-1}$) in the ultra strong UV FeII emitter Q2226-3905
 (Graham, Clowes, \& Campusano, 1996).

 All these similarities point towards NLS1s being low redshift, low
 luminosity analogues of high redshift quasars.

\section[]{Do NLS1s reside in rejuvenated galaxies?}

 In the previous section we have argued that NLS1s may represent an
 early phase in AGN evolution. Whether they reside in young galaxies
 is a separate question and a step further. That young galaxies are
 gas rich is helpful; they would have the large reservoir of gas
 necessary to sustain the close to Eddington rate accretion in
 NLS1s. But do we have any evidence that they indeed reside in young
 galaxies? There is no published systematic study of the properties of
 the host galaxies of NLS1s. However, some of the NLS1s are originally
 from Zwicky (e.g. I Zw 1) and Markarian (e.g. Mrk 766) sample of
 galaxies implying that they are blue. While the blue color might be
 due to big blue bumps in the active nuclei, as in normal Seyfert
 galaxies, NLS1s have weak blue bumps ($\S 1$) and so the blue colors
 might be a result of actively star forming galaxies. Some NLS1s are
 IRAS galaxies (e.g. IRAS 13349+2438), infra-red bright, and star
 forming.  Using the catalogs of galaxies RC3 (de Vaucouleurs
 \etal~1991) and UGC (Nilson, 1973) we looked into the morphology of a
 small sample of NLS1s with X-ray absorption features and found
 information on seven of them. Three were found to be compact (I Zw 1,
 Mrk 507, and Mrk 1298), two with signatures of inner ring (NGC 4051
 and Ark 564), and three with nuclear bars (NGC 4051, Mrk 776 and Ark
 564).  These are signatures of recent activity, most likely due to
 galaxy- galaxy interactions or mergers.  In this scenario the
 galaxies are newly formed, or rejuvanated.

 That NLS1s reside in young galaxies is also consistent with the
 hypothesis that the formation and evolution of galaxies and their
 active nuclei is intimately related (Rees 1997, Fabian 1999, Granato
 \etal 1999, Haehnett \& Koffmann 1999). In this scenario, the process
 of formation of a massive BH and the active nucleus is the very
 process of galaxy formation. The active nucleus and the galaxy evolve
 together, with BH accreting matter and the galaxy making stars. At
 one stage the winds from the active nucleus blow away the matter
 surrounding it and a quasar emerges. This is not only the end of
 active evolution of the quasar but that of the galaxy as well, which
 is evacuated of its interstellar medium. The quasar then shines as
 long as there is fuel in the accretion disk (Fabian 1999). In this
 scenario, high redshift quasars represent early stage of galaxy
 evolution, BALQSOs at z$\approx 2$ represent the stage when the gas
 is being blown away and z$\approx 1$ quasars would be the passively
 evolving population. Massive ellipticals found today might be the
 dead remnants of what were once quasars.

 The quasar phenomenon may thus be a result of galaxy formation due to
 primordial density fluctuations. At low redshift, when new galaxies
 are formed due to interactions or mergers, similar evolution may take
 place. As argued above, the NLS1s may represent a crucial early
 phase. (In our scenario, the accretion rate \.m is large in the early
 stages of evolution and reduces later on. This is opposite to the
 proposal by Wandel (1999) in which \.m increases with time.)

 In fact, there might be some NLS1s with a star burst component (see
 $\S 2$). Soft X-ray spectra of NLS1 are steep and often
 variable. However, Page \etal~(1999) report that while the power-law
 component in the NLS1 Mrk 766 varied, the thermal black-body
 component did not. This component might well be due to a nuclear
 starburst. Note also the strong CO emission in the prototype NLS1 I Zw
 1 (Barvainis, Alloin \& Antonucci 1989). Schinnerer, Eckart \&
 Tacconi (1998) mapped I Zw 1 in Co and found a circumnuclear ring of
 diameter 1.8 kpc. The authors found strong evidence for a nuclear
 starburst. There is also a companion to I Zw 1, supporting an interpretation
 of starburst  due to interaction. Similarly, AGN activity
 is known to exist in star burst galaxies (see Heckman 1999 for a
 review). Dennefeld \etal~(1999) report observations of narrow optical
 emission lines in a sample of IR selected starburst galaxies.

\section{Observational Tests}

 Here we propose several observations that could test the ideas
 presented above. (1) Emission line properties of high redshift
 quasars: Objects in the Shields \& Hamann sample, for example, were
 selected on the basis of their colors, in particular very red B$-$R
 which results when Ly$\alpha$ is redshifted into the R band. This
 could lead to a selection bias in favor of objects with narrow, peaky
 profiles (Shields \& Hamann 1997). It would be important to remove
 such selection bias before we can conclude that z$>4$ quasars have
 narrow emission lines. A broader wavelength coverage with more than
 just two bands would be useful. The ``drop-out'' technique (Steidel
 \etal 1995) used for finding Lyman break galaxies would be another
 way to remove emission line bias. (2) X-ray properties of high
 redshift quasar: Only about a dozen z$>4$ quasars are detected in
 X-rays (see the latest update by Kaspi, Brandt \& Schneider,
 2000). However, X-ray spectra of z$>4$ quasars are still not
 available. It will be interesting to see if they are steep, and
 highly variable like those of NLS1s. We will be studying X-ray
 spectra of radio-loud as well as radio quiet z$> 4$ quasars with XMM.
 (3) Morphology and environment of NLS1s: Are NLS1s preferentially
 found in younger galaxies and/or in more disturbed environments?  A
 systematic study of host galaxy properties of a well defined sample
 of NLS1s is needed.  (4) Search for starburst components in NLS1s:
 Evidence for the starburst--NLS1 connection ($\S 4$) is suggestive,
 but not yet statistically sound. Do the nuclear components of
 starburst galaxies show narrow emission lines? Do NLS1s show evidence
 of a starburst component more often than normal Seyferts?
 Spectroscopic as well as high resolution imaging observations would
 help towards establishing the connection between the two.

\section[]{On the correlation of FeII strength and X-ray spectral slope}

 While photoionization models reproduce the properties of optical and
 UV emission lines observed in AGN spectra with reasonable success,
 the case of FeII lines is different. Wills, Netzer \& Wills (1985)
 found that standard photoionization models cannot explain the
 strength of the the observed FeII lines in AGN.  Collin-Souffrin and
 collaborators (1988) as well as Kwan (1984) advocated that
 collisional ionization would be important in the production of FeII.
 Collin (1999) has again shown that strong FeII emission cannot be
 produced by photoionization with any set of parameters, and even by
 making iron abundance reasonably super-solar.  The importance and
 necessity of collisional ionization of iron was, however, not
 appreciated at least in part due to the observed correlation between
 FeII$\lambda 4570$ equivalent width and the soft X-ray energy index
 (Wilkes, Elvis \& McHardy, 1987, Shastri \etal~1993). Standard
 photoionization models for the line emission from the broad line
 regions of quasars (Krolik 1988) imply a close link between the
 energy index of the ionizing X-ray continuum and the strength of
 emission lines. So the Wilkes \etal~ and Shastri \etal~ correlation
 was interpreted as a result of photoionization.  Note, however, that
 the correlation is in in the opposite sense to that predicted by the
 standard photoionization model in which FeII emission is generated
 deep within the cloud and thus is sensitive to harder X-rays. NLS1s
 may provide us with the clue to understand this observed, conflicting
 trend as discussed below.

 As discussed in $\S 1$, there is a general consensus that large
 accretion rate, \.M/\.M$_{Edd}$, is the likely driver of the many
 observed properties of NLS1s.  Large strength of FeII emission may
 then be linked to the large accretion rate. In a model by
 Kwan \etal~(1995) FeII line emission is produced in
 an accretion disk.
  The accretion disks with larger accretion rate may simply
 have more mass to produce stronger FeII. Thus we argue that the
 observed correlation of FeII strength and soft X-ray slope is a
 consequence of the correlation with the accretion rate and support
 collisional ionization as the origin.

\section[]{Conclusions}

 We have argued that NLS1s are likely to be AGN in the making and
 reside in rejuvenated galaxies. As such they represent a crucial
 early phase in the evolution of galaxies and active galaxies.  What
 we need now is a systematic study of host galaxy properties of a well
 defined sample of NLS1s and their comparison with the hosts of normal
 Seyferts. Some evidence presented above is based on a small number of
 objects and generalization may not be appropriate.  Studies at high
 redshift also suffer from selection effects. Understanding and
 correcting for them is crucial in establishing the analogy with NLS1s
 on firm footing. X-ray spectra of high redshift radio-quiet quasars
 will provide an addition piece of information towards this goal.
 It would of great interest to find out
 whether star burst galaxies are parent population of NLS1s.

\section*{Acknowledgments}

 I thought about NLS1s and their place in the cosmic ``big picture''
 while preparing my talk for the NLS1 workshop at Bad Honnef,
 Germany. I am grateful to Th. Boller and the organizing committee for
 inviting me to the workshop ``Observational and theoretical progress
 in the study of Narrow Line Seyfert 1 Galaxies''.

 It is my pleasure to thank F. Hamann, C. Reynolds, D. Weinberg,
 A. Pradhan, R. Pogge, B. Peterson, M. Elvis, B. Ryden and P. Osmer
 for useful discussions and encouragement, and the referee Niel Brandt
 for useful comments.

      This research has made use of the NASA/IPAC Extragalactic
      Database (NED) which is operated by the Jet Propulsion
      Laboratory, California Institute of Technology, under contract
      with the National Aeronautics and Space Administration. Support
      through NASA grant NAG5-3249 (LTSA) is gratefully acknowledged.

\bsp

\label{lastpage}

\end{document}